\documentclass[reprint, aip, two column, superscriptaddress, showkeys]{revtex4-1}
\usepackage{graphicx,epsfig}
\usepackage{amssymb}
\usepackage{amsmath}
\usepackage{bm}
\usepackage{textcomp}
\usepackage{color}

\begin{document}
\setcounter{page}{1}

\title[]{Tuning of Fermi Contour Anisotropy in GaAs (001) 2D Holes via Strain }
\author{Insun \surname{Jo}}
\author{M. A. \surname{Mueed}}
\author{L. N. \surname{Pfeiffer}}
\author{K. W. \surname{West}}
\author{K. W. \surname{Baldwin}}
\affiliation{Department of Electrical Engineering, Princeton University, Princeton, NJ 08544 USA}
\author{R. \surname{Winkler}}
\affiliation{Department of Physics, Northern Illinois University, DeKalb, IL 60115 USA}
\author{Medini \surname{Padmanabhan}}
\affiliation{Physical Sciences Department, Rhode Island College, Providence, RI 02908 USA}
\author{M. \surname{Shayegan}}
\affiliation{Department of Electrical Engineering, Princeton University, Princeton, NJ 08544 USA}
\date{\today}

\begin{abstract}
We demonstrate tuning of the Fermi contour anisotropy of two-dimensional (2D) holes in a symmetric GaAs (001) quantum well via the application of in-plane strain. The ballistic transport of high-mobility hole carriers allows us to measure the Fermi wavevector of 2D holes via commensurability oscillations as a function of strain. Our results show that a small amount of in-plane strain, on the order of $10^{-4}$, can induce significant Fermi wavevector anisotropy as large as 3.3, equivalent to a mass anisotropy of 11 in a parabolic band. Our method to tune the anisotropy \textit{in situ} provides a platform to study the role of anisotropy on phenomena  such as the fractional quantum Hall effect and composite fermions in interacting 2D systems. 
\end{abstract} 

\maketitle   

High quality two-dimensional (2D) holes in GaAs exhibit various quantum mechanical phenomena, rendering them an attractive platform for fundamental resarch as well as applications for novel electronic and spintronic devices. For example, the strong spin-orbit interaction in GaAs leads to the intrinsic spin Hall effect, \cite{Wunderlich.PRL.2005,Bernevig.PRL.2005} and 2D holes in narrow quantum wells (QWs) have shown long spin coherence times, making them promising cadidates for spin qubits. \cite{Korn.NJP.2010} Also, high-mobility hole carriers and their zero-field spin-splitting, tunable by external electric field, \cite{Heremans.SC.1994,Lu.PRL.1998,Papadakis.Science.1999,Papadakis.PhysE.2001,Fischer.PRB.2007} are essential components for mesoscopic spin devices. \cite{Heremans.SC.1994,Lu.PRL.1998,Rokhinson.PRL.2004,Nichele.PRL.2015} In addition, recent magnetotransport measurements reveal that the inverted (``Mexican hat'') dispersion of the hole excited subband hosts an exotic annular Fermi sea. \cite{Jo.PRB.2017} 

\begin{figure} [b!]
  \begin{center}
    \psfig{file=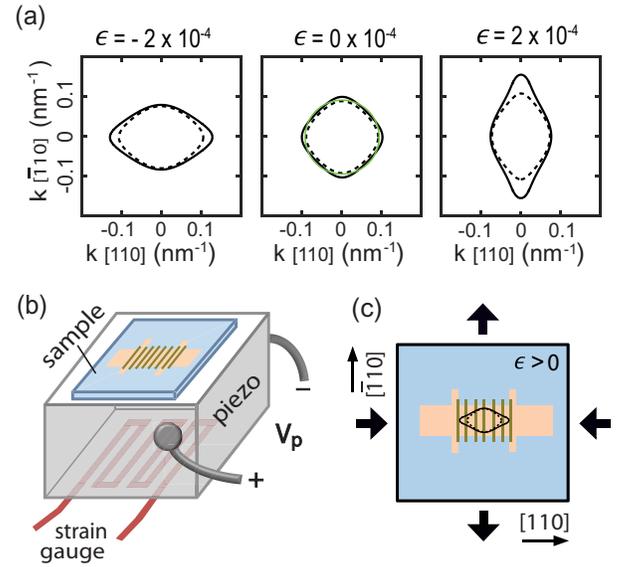, width=0.46\textwidth }
  \end{center}
  \caption{\label{fig1} (a) Calculated Fermi contours of 2D holes in a symmetric GaAs quantum well at density $p=1.3 \times 10^{11}$ cm$^{-2}$ in the presence of strain $\epsilon$ along the [$\bar{1}$10] direction. Solid and dashed contours represent two spin-split subbands; the green circle with radius $\sqrt{2 \pi p}$ shows a spin-degenerate, circular Fermi contour at the same density. (b) Schematic of the experimental setup, showing a thinned GaAs sample glued on a piezo-actuator. A voltage bias $V_\mathrm{P}$ applied to the piezo exerts strain along [$\bar{1}$10], and the resulting strain change is monitored by the strain gauge mounted underneath. (c) Sample etched into the Hall bar geometry has a periodic grating on the surface made of negative electron-beam resist. As a result, the magnetoresistance traces yield commensurability oscillations, which provide the Fermi wavevectors. Thick arrows indicate the deformation of the crystal for the case of $\epsilon >0$, and the resulting real-space cyclotron orbits are shown by black curves; note that these are rotated by $90^{\circ}$ with respect to the Fermi contour in reciprocal space. }
\end{figure} 

\begin{figure*}[t!]
  \begin{center}
    \psfig{file=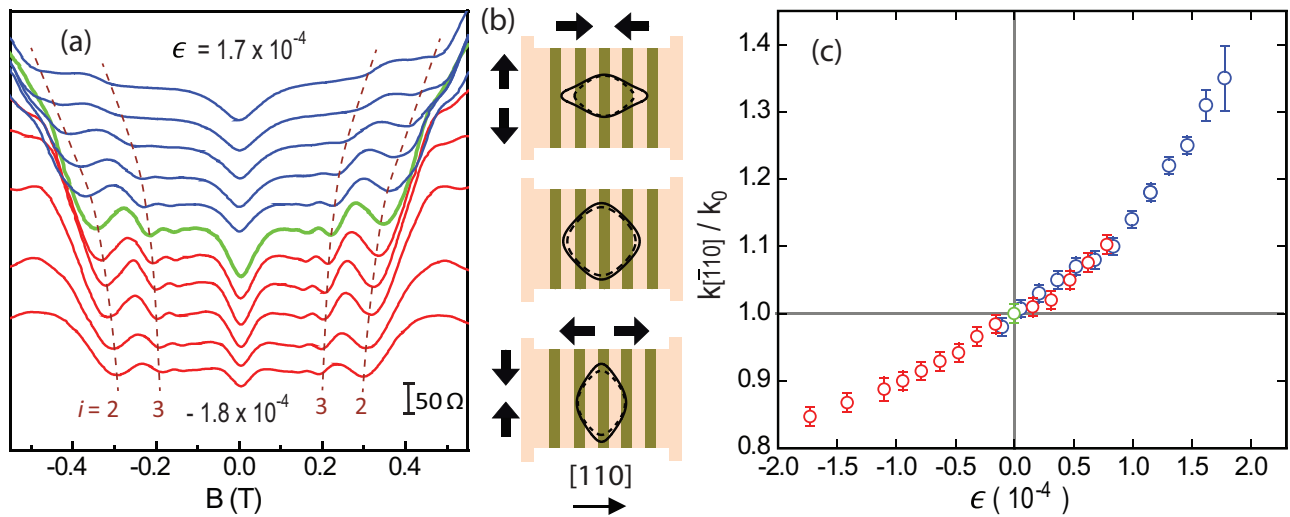, width=0.95\textwidth }
  \end{center}
  \caption{\label{fig2} (a) Magnetoresistance traces taken at $T= 0.3$ K from the Hall bar under in-plane strain $\epsilon$ between $- 1.8 $ and $1.7 \times 10^{-4}$. The green trace represents the strain-free case, $\epsilon =0$, while red (blue) traces are for the compressible (tensile) strain case. The commensurability minima appear symmetrically near $B = 0 $ and their evolutions with varying $\epsilon$ are marked by brown dashed lines. (b) The \textit{real-space} cyclotron orbits are depicted by black curves for different strain ranges: $\epsilon > 0$ (top), $\epsilon = 0$ (middle), $\epsilon < 0$ (bottom). The two cyclotron orbits shown by solid and dashed curves correspond to the two spin-split subbands. In the experiments, we detect only one orbit. Thick arrows indicate the directions of the crystal deformation as a result of strain. (c) The measured $k_{[\bar{1}10]}$, normalized to $k_0$, is shown as a function of $\epsilon$. Red and green circles are measured in the same cool-down while blue circles are measured at different cool-down. The magnetoresistance traces shown in (a) use the same color code. }
\end{figure*}

The band-structure modifications resulting from the application of in-plane strain to GaAs 2D holes also provide exciting opportunities for fundamental studies and device applications. Previous studies demonstrated the tuning of the spin-splitting and the Fermi contour shape by applying in-plane strain to the GaAs (311)\textit{A} 2D holes. \cite{Habib.PRB.2007,Shabani.PRL.2008,Habib.SST.2009} However, in such samples, despite the nearly isotropic Fermi contour, low symmetries in the wafer orientation and the interface corrugations cause severe anisotropic mobilities along the $[01\bar{1}]$ and $[\bar{2}33]$ directions. \cite{Heremans.JAP.1994,Wassermeier.PRB.1995} These complications can be eliminated by growing the 2D holes on a GaAs (001) substrate. There have indeed been magnetotransport studies on GaAs (001) 2D holes in the presence of strain, and these have reported an effective mass change for one of the spin-split subbands \cite{Kolokolov.PRB.1999} as well as a control of the interaction-induced anisotropic phase in high magnetic fields. \cite{Koduvayur.PRL.2011} Yet, \textit{quantitative} measurements of strain-induced Fermi contour anisotropy have not been reported.  In this Letter, we demonstrate tunable Fermi contour anisotropy via the application of in-plane strain to GaAs (001) 2D holes, and measure the Fermi wavevector anisotropy as a function of strain using commensurability oscillations.
  
Figure 1 highlights our study to induce anisotropic Fermi contours in 2D holes confined to a symmetric GaAs QW grown on a GaAs (001) substrate. In Fig. 1(a) we show the calculated Fermi contours at a density of $p=1.3 \times 10^{11}$ cm$^{-2}$ under various strain values applied along the $[\bar{1}10]$ direction. \cite{fnote1,Shayegan.APL.2003,Shkolnikov.APL.2004} The self-consistent numerical calculations are performed based on the $8\times8$ Kane model augmented by the Bir-Pikus strain Hamiltonian. \cite{Bir.Book,Winkler.Book} Owing to the bulk inversion asymmetry of the zinc-blende structure of GaAs, a finite spin-splitting is expected in our symmetric QW, as indicated by two Fermi contours (solid and dashed curves). \cite{fnote2} Without strain ($\epsilon=0$), the Fermi contours show a four-fold symmetry in reciprocal space; the smaller contour is close to circular and the larger contour is slightly warped. When tensile strain ($\epsilon >0$) is applied along $[\bar{1}10]$, the Fermi contours become elongated along $[\bar{1}10]$. For compressive strain ($\epsilon <0$), the distortion of the Fermi contours occurs in the opposite direction. Remarkably, the induced Fermi contour anisotropy is quite large even for a small amount of strain, of the order of $10^{-4}$. This pronounced distortion is attributed to the large hole effective mass and the strong heavy-hole and light-hole mixing in the band structure of GaAs 2D holes; \cite{Kolokolov.PRB.1999,Platero.PRB.1987,Winkler.Book,Sun.JAP.2007} the strain effect in GaAs 2D electrons is negligible. \cite{fnote3}

Figures 1(b) and 1(c) show the schematic of our experimental setup. Our sample contains 2D holes confined to a 175-{{\AA}}-wide, modulation-doped, GaAs QW grown on a GaAs (001) substrate by molecular beam epitaxy. The symmetric QW is flanked by a 960-{{\AA}}-thick Al$_{0.24}$Ga$_{0.76}$As spacer layer and a C $\delta$-layer on each side, resulting in a density $p \simeq1.3 \times 10^{11}$ cm$^{-2}$ and mobility $2 \times 10^6$ cm$^2$/Vs at 0.3 K. A $4 \times 4$ mm$^2$ piece of the wafer is thinned to $\sim 120~\mu$m using mechanical lapping followed by chemical-mechanical polishing. \cite{fnote4} A Ti/Au gate is deposited on the back-side of the wafer to tune the density and also to shield any spurious external field from the piezo-actuator. The thinned sample, etched into the Hall bar geometry, is glued on one side of the stacked piezo-actuator using two-component epoxy, and cooled to 0.3 K for the magnetotransport measurements. A strain gauge is also glued on the other face of the piezo to monitor the relative change of the strain as a voltage bias $V_\mathrm{P}$ is applied to the piezo. \cite{Shayegan.APL.2003,Shkolnikov.APL.2004} After sample cool-down, a finite built-in strain develops due to the different thermal contractions of the sample and the piezo. We determine the $\epsilon =0$ condition when the measured Fermi wavevector equals $k_0=\sqrt{2 \pi p}$. (see Ref. 29.)

In order to measure the Fermi wavevector, we fabricate a grating of negative electron-beam resist with period $a= 200$~nm on the surface of the wafer. The grating induces a periodic strain on the GaAs surface which in turn results in a small periodic modulation of the 2DHS density via the piezoelectric effect . \cite{Endo.PRB.2000, Kamburov.PRB.2012a} When a small perpendicular magnetic field $B$ is applied, the holes move along the cyclotron orbits, whose shapes resemble the Fermi contours, but rotated by $90^{\circ}$. When the cyclotron diameter along the current direction becomes commensurate with $a$, the magnetoresistance exhibits commensurability oscillations, also known as Weiss oscillations. \cite{Weiss.EPL.1989,Winkler.PRL.1989,Gerhardts.PRL.1989,Beenakker.PRL.1989} The minima in the trace are directly related to the 2D holes' Fermi wavevector along $[\bar{1}10]$, $k_{[\bar{1}10]}$, through $ 2R_c /a = i-1/4$ ($i=1,2,3 ...$) where 2$R_{c} = 2\hbar k_{[\bar{1}10]} /eB$ is the cyclotron diameter along $[110]$, $e$ is the electron charge, and $\hbar$ is the Planck constant. \cite{Weiss.EPL.1989,Winkler.PRL.1989,Gerhardts.PRL.1989,Beenakker.PRL.1989,Endo.PRB.2000,Kamburov.PRB.2012a} [As an example, the shapes of the cyclotron orbits for $\epsilon >0$ are depicted by black curves in Fig. 1(c).]	
 
Figure 2(a) shows the measured magnetoresistance traces at different $\epsilon$. The green trace represents the $\epsilon =0$ case, satisfying $k_{[\bar{1}10]}=k_0$. The blue traces are for tensile strain ($\epsilon >0$) while the red ones are for compressive strain ($\epsilon <0$). The commensurability features appear symmetrically with respect to $B=0$, and the minima positions corresponding to $i=2$ and 3 are used to determine $k_{[\bar{1}10]}$. The $i=1$ minimum is expected at higher field ($\simeq 0.8$ T for $\epsilon=0$), but it is masked by the Shubnikov-de Haas oscillations. The evolution of the mimima positions as a function of strain is clearly seen in Fig. 2(a) following the dashed lines; with increasing $\epsilon$ the minima move away from $B=0$, implying an increasing $k_{[\bar{1}10]}$. In Fig. 2(b), black solid and dashed curves represent \textit{real-space} cyclotron orbits of holes in different $\epsilon$ regimes; $\epsilon >0$ (top), $\epsilon=0$ (middle), and $\epsilon<0$ (bottom). In principle, there are two cyclotron orbits corresponding to the two spin-subbands. However, we measure a single $k_{[\bar{1}10]}$ from the commensurability minima, implying that our experiments do not resolve the finite but small spin-splitting. The measured $k_{[\bar{1}10]}$ normalized to $k_0$ is shown in Fig. 2(c) as a function of $\epsilon$. The red and blue circles are obtained from two different cool-downs. Although the built-in strains are different between the two cool-downs, both sets of data contain $k_{[\bar{1}10]}=k_0$, i.e., $\epsilon=0$. This enables us to determine the built-in strain, and thereby the absolute values of $\epsilon$. It is clear that the overlapping $k_{[\bar{1}10]}$ values from two measurements agree very well each other. When $\epsilon$ is increased from $-1.8$ to $1.7 \times 10^{-4}$, $k_{[\bar{1}10]}/k_0$ changes from 0.85 to 1.38, i.e., there is a $\simeq 60\%$ increase of the Fermi wavevector along the $[\bar{1}10]$ direction.

\begin{figure}
  \begin{center}
    \psfig{file=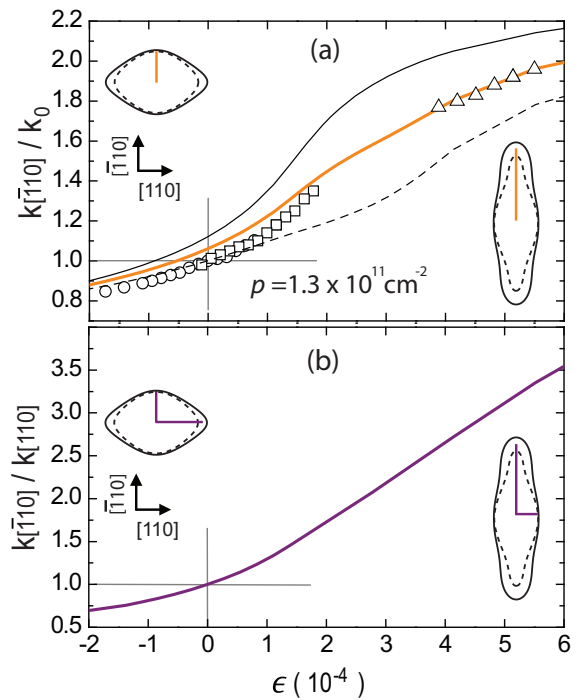, width=0.42\textwidth }
  \end{center}
  \caption{\label{fig3} (a) Measured $k_{[\bar{1}10]}/k_{0}$ for different cool-downs, and the calculation results. Circles, squares, and triangles represent the measurement results from different cool-downs. Black solid and dashed curves represent the calculated Fermi wavevectors for the two spin-split subbands; their average values are depicted by the thick orange curve. For the triangles, the built-in strain is estimated by matching the measured $k_{[\bar{1}10]}/k_{0}$ to the thick orange curve. (b) The calculated Fermi wavevector anisotropy, defined by $k_{[\bar{1}10]}/k_{[110]}$ is shown at different $\epsilon$. The left and right insets in (a) and (b) show the shapes of Fermi contours for $\epsilon=-2.0$ and $5.5 \times 10^{-4}$, respectively. }
\end{figure}

Figure 3(a) shows the measured $k_{[\bar{1}10]}/k_0$ and its comparison with the calculations as a function of $\epsilon$. The circles, squares, and triangles represent the measured $k_{[\bar{1}10]}/k_0$ in three different cool-downs where different built-in strains were attained. The black solid and dashed lines represent the calculated $k_{[\bar{1}10]}/k_0$ for the two spin-subbands, and the thick orange curve shows their averaged values. The data shown by circles and squares are slightly lower than the orange curve. The small discrepancy indicates that the measured $k_{[\bar{1}10]}$ deduced from the commensurability features may not reflect exactly the averaged $k_{[\bar{1}10]}$ of the two spin-subbands, or the calculated Fermi contours are slightly different from those in our sample. Compared to the circles and squares, the measured data plotted by triangles show very large $k_{[\bar{1}10]}/k_0$ values, implying that a large built-in strain develops for this cool-down. \cite{Shabani.PRL.2008,fnote5} Because $k_{[\bar{1}10]}$ cannot be tuned over a sufficiently large range to reach $k_0$ by applying $V_\mathrm{P}$ to the piezo, limited within $\pm~300$~V, we are unable to determine the absolute magnitude of $\epsilon$ for this cool-down directly from the measured data. If we assume a built-in strain of $\simeq 4.7 \times 10^{-4}$, however, the measured data match very well with the orange curve. Overall, the experimental data in Fig. 3(a) agree well with the averaged Fermi wavevector of the calculations to within~6\%. 

Figure 3(b) shows the Fermi wavevector anisotropy, defined by $k_{[\bar{1}10]}/k_{[110]}$ where we use the average values of the two spin-subbands for each direction, based on calculations results for $p=1.3 \times 10^{11}$ cm$^{-2}$. A remarkably large Fermi wavevector anisotropy, as high as 3.3, is achieved when $\epsilon= 5.5 \times 10^{-4}$ is applied. Note that this is equivalent to an effective mass anisotropy of 11 in a parabolic band, much larger than the intrinsic mass anisotropy ($\simeq 5$) of semiconductors such as Si and AlAs. \cite{Sze.Book,Shayegan.PSSb.2006} Moreover, the tunability of the anisotropy in 2D holes allows for systematic studies of anisotropy effect on transport properties, which is not viable with the fixed, anisotropic 2D electrons in Si and AlAs. 

Before closing, we note that recent experiments have demonstrated that parallel magnetic fields can induce anisotropic Fermi contours in high-quality GaAs 2D hole (and electron) systems and that the induced anisotropy can be determined via commensurability oscillations measurements. \cite{Kamburov.PRB.2012b,Kamburov.PRB.2013,Mueed.PRL.2015} However, the induced anisotropy is primarily caused by the coupling between the in-plane and out-of-plane motions of the carriers, making the theoretical understanding of the data challenging. In contrast, the strain-induced anisotropy is originated from the band structure modifications at zero magnetic field. We also emphasize several different aspects between the two methods. First, the in-plane strain can both expand and contract the Fermi contour along a particular direction, while parallel-field can only elongate the Fermi contour in the direction perpendicular to the applied field direction. Second, the out-of-plane motions of the carriers driven by the in-plane field is affected by the QW width, thus the induced anisotropy is strongly depedent on the QW width as well as the out-of-plane effective mass, \cite{Kamburov.PRB.2014} while the strain does not bring in such complexity to the 2D holes. Third, a large parallel field inevitably leads to a large Zeeman spin-splitting of 2D holes. To this end, 2D holes exhibiting \textit{tunable} anisotropy under strain provide an ideal platform to study the role of anisotropy in phenomena such as the fractional quantum Hall effect and composite fermions in interacting 2D systems. \cite{Kamburov.PRL.2013,Kamburov.PRB.2014,Mueed.PRB.2016,Balagurov.PRB.2000,Gokmen.NatPhys.2010,Haldane.PRL.2011,KYang.PRB.2013,Mueed.PRB.2016,Balram.PRB.2016,Jo.arXiv.2017,Ippoliti.arXiv.2017}

\begin{acknowledgments}
  We acknowledge support by the NSF Grant ECCS 1508925 for measurements, and Grants by the DOE BES (DE-FG02-00-ER45841), the NSF (DMR 1305691 and MRSEC DMR 1420541), and the Gordon and Betty Moore Foundation (Grant GBMF4420) for sample fabrication and characterization. Our theoretical work was supported by the NSF Grant DMR 1310199.
\end{acknowledgments}


\begin{thebibliography}{99}
\bibitem{fnote2} The QW used in this study is symmetrically doped, thus the average electric field in the QW is close to zero. The calculations are performed for the zero electric field case.
\bibitem{fnote3} The electrons in the conduction band inherit the strain-induced anisotropy from the states in the valence band.  For the electrons, this is then a higher-order effect than the anisotropy of the holes.
\bibitem{fnote4} We used Logitech PM5 polisher/lapper to thin the sample. The front side of the sample is protected by PMMA electron-beam resist and glued to a glass plate with crystal bond. 2-$\mu$m aluminum oxide grits are used for the initial lapping, and Chemlox SF-1 suspension fluid is used for the final polishing.
\bibitem{fnote5} Although the built-in strain for each cool-down is not completely controllable, we achieved a relatively small built-in strain when the sample was cooled with $V_\mathrm{P}$ open-circuitted (empty circles and squares), while large built-in strain was induced (empty triangles) when a 1 G$\Omega$ resistor was attached to the two leads of the piezo-actuator during the cool-down.
\end{thebibliography}
\end{document}